\documentclass[submission,copyright,creativecommons]{eptcs}
\providecommand{\event}{CREST 2018} 
\usepackage{breakurl}             
\usepackage{underscore}           

\usepackage{amsthm}  

\usepackage{amsmath}
\usepackage{latexsym}
\usepackage{amssymb}
\usepackage{verbatim}
\usepackage{setspace}
\usepackage{enumerate}

\usepackage{graphicx}
\usepackage{tikz}
\usetikzlibrary{tikzmark}

\usepackage{tikz}
\usetikzlibrary{arrows,decorations.pathmorphing,backgrounds,positioning,fit}

\newtheorem{df}{Def}[section]  
\newtheorem{teo}[df]{Theorem}

\newtheorem{example}[df]{Example}

\newtheorem{notat}[df]{Notation}

\newcommand{\dep}[2]{=\hspace{-3pt}({#1};{#2})}

\newcommand{\cf}{\hspace{2pt}\Box\hspace{-4pt}\rightarrow}

\newcommand{\CD}{\mathcal{CD}}

\newcommand{\PCD}{\mathcal{PCD}}

\newcommand{\CO}{\mathcal{CO}}

\newcommand{\SET}[1]{\mathbf{#1}}

\author{Fausto Barbero and Gabriel Sandu (University of Helsinki)}
\title{Interventionist Counterfactuals on Causal Teams}
\def\titlerunning{Interventionist Counterfactuals on Causal Teams}
\def\authorrunning{Fausto Barbero and Gabriel Sandu}
\date{}

\begin{document}

\setcounter{page}{1} 

\maketitle

\begin{abstract}
We introduce an extension of team semantics (\cite{Hod1997}, \cite{Vaa2007}) which provides a framework for the logic of  
manipulationist theories of causation based on structural equation models, such as 
Woodward's (\cite{Woo2003}) and Pearl's (\cite{Pea2000}); our causal teams incorporate (partial or total) information about 
functional dependencies that are invariant under interventions. We give a unified treatment of  
observational and causal aspects of causal models by isolating two operators on causal 
teams which correspond, respectively, to conditioning and to interventionist counterfactual 
implication.  We then introduce formal languages for deterministic and probabilistic causal discourse, and show how various notions of cause (e.g. direct and total causes) may be defined in them. 

Through the tuning of various constraints on structural equations (recursivity, existence and uniqueness of solutions, full or partial definition of the functions), our framework can  capture  different causal models. We give an overview of the inferential aspects of the recursive, fully defined case; and we dedicate some attention to the recursive, partially defined case, which involves 
a shift of attention towards nonclassical truth values.
\end{abstract}

\section{Introduction}

Some modern accounts of causation, most eminently the framework of D. Lewis (\cite{Lew1973b}), link the notion of causality to that
of counterfactual dependence.
Recent approaches to the manipulationist
analysis of causality (Pearl \cite{Pea2000}, Spirtes, Glymour \& Scheines \cite{SpiGlySch1993}, Woodward \cite{Woo2003}, Hitchcock \cite{Hit2001})
focus on counterfactuals whose antecedents express \emph{interventions}, the key idea being that a cause can be intervened upon to determine its effect. Such theories articulate the analysis of the notion of intervention using the so-called \textit{structural equation models} (\cite{SpiGlySch1993}, \cite{Pea2000}); they will be our main concern in the present paper. Our goal is to
show how the notions of counterfactual and causal dependence that
arise from the manipulationist theories of causation can be expressed
and incorporated in the logical framework provided by \textit{team
semantics}. In section \ref{SECSEM}, we briefly review the basics of structural equation modeling. In section \ref{SECCT}, we review the notion of a team and show how to integrate it with causal structure. Sections \ref{BASICLAN} to \ref{SECINTERV} define the (causal) team semantics for (deterministic) atomic formulas, connectives, and operators corresponding to evidential and counterfactual reasoning. Section \ref{SECRECPAR} briefly explores the properties of this language when evaluated in the context of recursive, fully defined systems; sections \ref{SECMOREGEN} and \ref{SECFALADM} sketch ideas for going beyond these restrictions. In section \ref{SECPROB} we show how to enrich the languages with probabilistic statements. Finally, as an example, we show that our languages are adequate for expressing the notions of direct and total causation from Woodward (\cite{Woo2003}). The reader can find a more extensive treatment of our subject (including the omitted proofs) in the preprint \cite{BarSan2017}.

\section{Structural equation models} \label{SECSEM}

The most basic objects in the structural equation modeling approach are \emph{variables}, which we will denote with capital letters $X,Y...$. Each variable $V$ can assume values (tipically denoted as $v,v',v''$...) within a certain range of objects, $Ran(V)$. Variables are related to each other by  \emph{structural equations}, for example 

\vspace{-10pt}

\[
Y := f_Y(X_1,\dots,X_n)
\]
stating that $Y$ is determined as a function of $X_1,\dots,X_n$. The use of the symbol $:=$ instead of an equality is to emphasize that the equation should be thought of as non-reversible\footnote{A structural equation is nothing else than a shorthand for a set of counterfactuals, to be taken as assumptions (\cite{Hit2001}).}. The set of arguments of function $f_Y$, that is $\{X_1,\dots,X_n\}$, is usually denoted as $PA_Y$ (the set of \emph{parents} of $Y$; $Y$ is a \emph{child} of each of the $X_i$). For other sets or sequences of variables, we will follow a different notational convention:

\begin{notat}
 \begin{itemize}
\item We use boldface letters such as \textbf{X} to denote either a set $\{X_1,\dots,X_n\}$ of variables or a sequence of the same variables (in a fixed alphabetical order). 
\item We use \textbf{x} to denote a set or sequence of values, each of which is a value for exactly one of the variables in \textbf{X}. We leave the details of these correspondences between variables and values as non-formalized.  
\item $Ran(\SET X)$ is an abbreviation for $\prod_{X\in \SET X} Ran(X)$.
\end{itemize}
\end{notat}

A structural equation model may contain an explicit description of the function $f$ (\emph{fully defined} case) or not (\emph{partially defined} case). In both cases, the structural equations determine a pattern of dependencies between variables, which can be represented as a graph (one arrow from each parent $X_i$ to the  child $Y$). 

An \emph{intervention} $do(X=x)$ can be thought of as the act of replacing the equation for $X$ with a constant equation $X:=x$. Correspondingly, all the arrows coming into $X$ are removed from the graph. Importantly, all the other structural equations are left untouched by the intervention. This aspect of the system of structural equations, called invariance (modularity) will be crucial in our developments.

A structural equation model is typically further enriched, in the literature, with an assignment of values to the exogenous variables (deterministic case), or with a joint probability distribution over  the exogenous variables (semi-deterministic case). If the graph underlying the model is acyclic, this assignment or probability distribution can be canonically extended to the whole variable domain. At this stage it becomes possible to evaluate counterfactual statements over the model: for example, $X=x\cf \psi$ holds under the current assignment/probability distribution if $\psi$ holds after the intervention $do(X=x)$.

\section{Causal teams} \label{SECCT}

Team semantics was introduced by Hodges (\cite{Hod1997})
 to provide a compositional presentation of the (game-theoretically defined) semantics of Independence-Friendly logic (\cite{HinSan1989},\cite{ManSanSev2011}). In the subsequent years, team semantics has been used to extend first-order logic by database dependencies (e.g. \cite{Vaa2007}, \cite{GraVaa2013}, \cite{Gal2012}); and to enrich propositional logics (e.g. \cite{YanVaa2016})
 and modal logics (\cite{Vaa2008}). Appropriate generalizations  have been used as descriptive languages for probabilistic dependencies (\cite{DurHanKonMeiVir2016}), quantum phenomena (\cite{HytPaoVaa2015}), Bayes nets (\cite{CorHytKonPenVaa2016}). 

The basic idea of team semantics is that notions such as dependence and independence, which express properties of relations, cannot be captured by Tarskian semantics, which evaluates formulas on single assignments\footnote{This can be formally proved, see \cite{CamHod2001}.}; the appropriate unit for semantical evaluation is instead the \emph{team}, i.e., a \emph{set} of assignments (all sharing a common variable domain). In our context, an assignment can be thought of as a way to encode a possible configuration for the values of variables; once a set $Dom$ of variables is fixed, each assignment will be a function $s:Dom \rightarrow \bigcup_{X\in Dom}Ran(X)$ such that $s(X)\in Ran(X)$ for each $X\in Dom$ (in the statistical literature, $s$ would be called an \emph{individual}). A \emph{team} $T$ of domain $dom(T) = Dom$ is a set of such assignments.

A significant example of a property that can be satisfied by a team is functional dependence (among variables). The formula $\dep{\SET X}{Y}$, called a dependence atom, has the intended meaning: the (values) of the variable Y are functionally determined by (the values) of the set of variables $\SET X$. Its truth in a team $T$ is defined by the following clause: 

\vspace{-10pt}

 \[
T\models\dep{\SET X}{Y} \iff \text{for all } s,s'\in T, \text{ if } s(\SET X) = s'(\SET X) \text{ then } s(Y) = s'(Y) 
\]
where $s(\SET X)= s'(\SET X)$ is an abbreviation for ``$s(X_1)=s'(X_1)$ and... and $s(X_n)=s'(X_n)$''. 

Teams turned out to be a very useful framework for describing data-driven correlations. But they are not sufficient, by themselves, to handle causal dependencies. The latter require that the functional correlations be robust, i.e. invariant under interventions. We thus need to enrich teams with a set of functions, the \emph{invariant functions}, which are the carriers of causal dependencies\footnote{The invariant functions will univoquely associate a set of structural equations to the enriched team.}; and we need to formulate the notion of intervention. We now move to technicalities.


Given a team $T^-$ and  $X\in dom(T^-)$, we write $T^-(X)$ for the set of values that are obtained for $X$ in the team $T^-$: $T^-(X):= \{s(X)|s\in T^-\}$. As before, we say that $T^-$ satisfies a dependence atom $\dep{\SET X}{Y}$, and we write $T^-\models \dep{\SET X}{Y}$, if, whenever $s(\SET X) = s'(\SET X)$ for all $s,s'\in T^-$, we have $s(Y) = s'(Y)$.

\begin{df}
A \textbf{causal team} $T$ over variable domain $dom(T)$ with endogenous variables $\mathbf V\subseteq dom(T)$ is a quadruple $T = (T^-,G(T),\mathcal{R}_T,\mathcal{F}_T)$, where:
\begin{enumerate}
\item $T^-$ is a team.
\item $G(T) =(dom(T),E)$ is a graph over the set of variables. For any $X\in dom(T)$, we denote as $PA_X$ the set of all variables $Y\in dom(T)$ such that the arrow $(Y,X)$ is in $E$.
\item $\mathcal{R}_T = \{(X,Ran(X))|X\in dom(T)\}$ 
 (where the $Ran(X)$ may be arbitrary sets) is a function which assigns a range to each variable
\item $\mathcal{F}_T$ is a function $\{(V_i,f_{V_i})|V_i\in\mathbf V\}$ that assigns to each endogenous variable a $|PA_{V_i}|$-ary function $f_{V_i}:dom(f_{V_i})\rightarrow 
ran(V_i)$ \hspace{20pt}(for some ${dom(f_{V_i})\subseteq Ran(PA_{V_i})}$) 
\end{enumerate}
which satisfies the further restrictions:
\begin{enumerate}[a)]
\item $T^-(X) \subseteq Ran(X)$ for each $X\in dom(T)$
\item If $PA_Y=\{X_1,\dots, X_n\}$, then $T^-\models \dep{X_1,\dots, X_n}{Y}$
\item if $s\in T^-$ is such that $s(PA_Y)\in dom(f_Y)$, then $s(Y)= f_Y(s(PA_Y))$.
\end{enumerate}
In case $dom(f_V)= Ran(PA_V)$ for each $V\in \SET V$, we say the causal team is \textbf{fully defined}; otherwise it is \textbf{partially defined}. If the graph $G(T)$ is acyclic, we say $T$ is \textbf{recursive}; otherwise \textbf{nonrecursive}. 

We will assume for the rest of the paper that $dom(T)$, and therefore $G(T)$, is finite.
\end{df}

Clause b) ensures that the team component $T^-$ satisfy (at least) the dependencies encoded in the graph $G(T)$. 
Clause c) further ensures that the team component is consistent with the invariant functions encoded in $\mathcal{F}_T$. The graph $G(T)$ 
is induced (via b) and c) ) by the set of functional dependencies specified by clause (4), and provides a distinction between \emph{endogenous} variables, that are determined by one of these invariant dependencies, and \emph{exogenous} variables (those in $dom(T)\setminus \SET V$), that are not. 

%
%

\begin{example} \label{EXCAUSALTEAM}
 Consider a causal team $T$ with underlying team $T^- =\{\{(U,2),(X,1),(Y,2),(Z,4)\},$ $\{(U,3),$ $(X,1),(Y,2),(Z,4)\},\{(U,1),(X,3),(Y,3),$ $(Z,1)\},\{(U,1),(X,4),(Y,1),(Z,1)\},\{(U,4),(X,4),$ $(Y,1),(Z,1)\}\}$, graph $G(T) = (\{U,$ $X,Y,Z\},$ $ \{(U,Z),(X,Y),(X,Z),(Y,Z)\})$, ranges $Ran(U) = Ran(X)$ $ = Ran(Y) = Ran(Z) = \{1,2,3,4\}$, and partial description of (one value of) the invariant function for $Z$: $\mathcal F(Z)(4,1,2):= 3$. We represent the $T^-$ and $G(T)$ components of $T$ by means of a decorated table:
\begin{center}
\begin{tabular}{|c|c|c|c|}
\hline
 \multicolumn{4}{|l|}{ } \\
 \multicolumn{4}{|l|}{U\tikzmark{U8} \ \ \tikzmark{X8}X\tikzmark{X8'} \ \ \ \tikzmark{Y8}Y\tikzmark{Y8'} \, \ \tikzmark{Z8}Z} \\
\hline
 2 & 1 & 2 & 4\\
\hline
 3 & 1 & 2 & 4\\
\hline
 1 & 3 & 3 & 1\\
\hline
 1 & 4 & 1 & 1\\
\hline
4 & 4 & 1 & 1\\ 
\hline
\end{tabular}
 \begin{tikzpicture}[overlay, remember picture, yshift=.25\baselineskip, shorten >=.5pt, shorten <=.5pt]
  \draw [->] ([yshift=3pt]{pic cs:X8'})  [line width=0.2mm] to ([yshift=3pt]{pic cs:Y8});
	\draw [->] ([yshift=3pt]{pic cs:Y8'})  [line width=0.2mm] to ([yshift=3pt]{pic cs:Z8});
  \draw ([yshift=7pt]{pic cs:X8'})  edge[line width=0.2mm, out=35,in=125,->] ([yshift=6pt]{pic cs:Z8});
	\draw ([yshift=8pt]{pic cs:U8})  edge[line width=0.2mm, out=35,in=125,->] ([yshift=8pt]{pic cs:Z8});
  \end{tikzpicture}
\end{center}
\end{example}

\section{A basic language for causal teams}   \label{BASICLAN}

We need first of all to specify what it means for a causal team to satisfy an atomic formula, and to assign a semantics to connectives. Our language consists of formulas built using the connectives $\land$ and $\lor$ (``tensor'' disjunction), dependence atoms, and atomic formulas of the forms  $Y=y$ and $Y\neq y$. The semantic clause for disjunction requires the notion of causal subteam:


\begin{df}
Given a causal team $T$, a \textbf{causal subteam} $S$ of $T$ is a causal team with the same domain and the same set of endogenous variables, which satisfies: 1) $S^-\subseteq T^-$, 2) $\mathcal{R}_S = \mathcal{R}_T$, 3) $G(S) = G(T)$,  4) $\mathcal{F}_S = \mathcal{F}_T$\footnote{Alternatively, one might consider enriching the component $\mathcal{F}_S$ with the information about invariant functions which is lost in passing from the team to the subteam.}.
\end{df}
The semantic clause for dependence atoms was given above. The other clauses are:
\begin{itemize}
\item $T\models Y=y$ (resp. $T\models Y\neq y$) if, for all $s\in T^-$, $s(Y)=y$ (resp. $s(Y)\neq y$) 
\item $T\models \psi\land \chi$ if $T\models \psi$ and $T\models \chi$.
\item $T\models \psi\lor \chi$ if there are causal subteams $T_1,T_2$ of $T$ s.t. $T_1^-\cup T_2^- = T^-$, $T_1\models \psi$ and $T_2\models \chi$.\footnote{Notice that it might be impossible to define consistently the union of two \emph{causal} teams.}
\end{itemize}

\section{Selective implication}

Our main goal is to give an exact semantics to counterfactual statements
of the form ``If $\psi$ had been the case, then $\chi$ would have
been the case''. Very often, however, one finds examples in the literature
where these statements are embedded into a larger context. 
Pearl (\cite{Pea2000}) analyzes the following query: ``what is the probability $Q$ that a subject who died under
treatment $(X=1,Y=1)$ would have recovered $(Y=0)$ had he or she
not been treated $(X=0)$? 

The representation of the statement whose probability Pearl is interested in seems to be:
\[
(X=1\wedge Y=1)\supset(X=0\cf Y=0).
\]
where the symbol $\cf$ stands for counterfactual implication, while the symbol $\supset$, called selective implication, denotes a connective which is a generalization of material implication to teams\footnote{To the best of our knowledge, this connective has been used, with different notation, in \cite{Gal2015}, as a special case of the \emph{maximal implication} introduced in \cite{KonNur2011}.}. It serves to restrict, in this example, the range of application of the counterfactual to the available evidence. 
More generally, 
given a causal team $T$, and a formula $\psi$ without dependence atoms, define the causal  subteam $T^\psi$ by the condition $(T^\psi)^- = \{s\in T^- | s\models \psi\}$,   
where the relationship $\models$ for single assignments is intended as in classical logic: $s\models Z=z$ if $s(Z) = z$, etc. We define selective implication by the clause:
\begin{itemize}
\item $T\models \psi
\supset \chi$ iff $T^\psi
\models \chi$.
\end{itemize}
The consequent $\chi$ can be any formula of the current language. Instead, we require the antecedent to be a formula which denotes properties of single assignments.
It is straightforward to extend the clause above in order to allow the use of $\supset$ (and the counterfactual $\cf$, yet to be defined) in antecedents. 

\begin{example}
The selective implication $Z=3 \supset Y=2$ 
holds on any causal team $T$ which has the table depicted below.
In order to see that the formula holds on it, we have to construct the subteam $T^{Z=3}$
\begin{center}
$T: $
\begin{tabular}{|c|c|c|}
\hline
 \multicolumn{3}{|c|}{Z \ Y \ X} \\
\hline
 1 & 2 & 3 \\
\hline
 2 & 1 & 1 \\
\hline
 3 & 2 & 1 \\
\hline
 3 & 2 & 2 \\
\hline
\end{tabular}
\hspace{20pt}$\leadsto$\hspace{20pt}
$T^{Z=3}: $
\begin{tabular}{|c|c|c|}
\hline
 \multicolumn{3}{|c|}{Z \ Y \ X} \\
\hline
 3 & 2 & 1 \\
\hline
 3 & 2 & 2 \\
\hline
\end{tabular}
\end{center}

which is obtained by selecting the third and fourth row of $T$ (the rows that satisfy $Z=3$). Notice that $T^{Z=3}\models Y=2$; by the semantical clause, then, $T\models Z=3 \supset Y=2$. 
 
\end{example}

\section{Intervention} \label{SECINTERV}
 We define an (interventionist) \emph{counterfactual implication}. Its semantics will be determined by a notion of \emph{intervention on a causal team}. We may think of a (causal) team as an incomplete description of our knowledge concerning the state of a system: each assignment represents a configuration of values for the variables that we consider possible, even though we do not know which specific assignment encodes the actual state of the system. If we perform an intervention on the system, say $do(X=1)$, then we know that, whatever the correct assignment is, our intervention is an action that enforces the values of the variable $X$ to take value $1$, and removes any causal link from other variables to $X$; it is then reasonable to apply these changes to the whole team. The change will then propagate to the descendants of $X$ by means of the functions specified by the fourth component of the causal team.


\begin{example} \label{EXINTERV}
Suppose we want to evaluate $X=1 \cf Y=2$ in the causal team of Example \ref{EXCAUSALTEAM}. We need to generate a causal team $T_{X=1}$ which differs from the initial one in that variable $X$ is fixed, in all assignments, to value 1. This will affect all descendants of $X$ (in this case, the children $Y$ and $Z$). 
\begin{center}
\begin{tabular}{|c|c|c|c|}
\hline
 \multicolumn{4}{|l|}{ } \\
 \multicolumn{4}{|l|}{U\tikzmark{U8bis} \  \  \tikzmark{X8bis}X\tikzmark{X8bis'} \ \  \  \tikzmark{Y8bis}Y\tikzmark{Y8bis'} \,  \ \  \tikzmark{Z8bis}Z} \\
\hline
 2 & 1 & 2 & 4\\
\hline
 3 & 1 & 2 & 4\\
\hline
 1 & 3 & 3 & 1\\
\hline
 1 & 4 & 1 & 1\\
\hline
4 & 4 & 1 & 1\\ 
\hline
\end{tabular}
 \begin{tikzpicture}[overlay, remember picture, yshift=.25\baselineskip, shorten >=.5pt, shorten <=.5pt]
  \draw [->] ([yshift=3pt]{pic cs:X8bis'})  [line width=0.2mm] to ([yshift=3pt]{pic cs:Y8bis});
	\draw [->] ([yshift=3pt]{pic cs:Y8bis'})  [line width=0.2mm] to ([yshift=3pt]{pic cs:Z8bis});
  \draw ([yshift=7pt]{pic cs:X8bis'})  edge[line width=0.2mm, out=35,in=125,->] ([yshift=6pt]{pic cs:Z8bis});
	\draw ([yshift=8pt]{pic cs:U8bis})  edge[line width=0.2mm, out=35,in=125,->] ([yshift=8pt]{pic cs:Z8bis});
  \end{tikzpicture}
\hspace{6pt}$\leadsto$\hspace{3pt}
\begin{tabular}{|c|c|c|c|}
\hline
 \multicolumn{4}{|l|}{ } \\
 \multicolumn{4}{|l|}{U\tikzmark{U9} \, \ \tikzmark{X9}X\tikzmark{X9'} \ \ \ \ \tikzmark{Y9}Y\tikzmark{Y9'} \ \ \ \  \  \tikzmark{Z9}Z} \\
\hline
 2 & \textbf{1} & $\dots$ & $\dots$\\
\hline
 3 & \textbf{1} & $\dots$ & $\dots$\\
\hline
 1 & \textbf{1} & $\dots$ & $\dots$\\
\hline
 1 & \textbf{1} & $\dots$ & $\dots$\\
\hline
4 & \textbf{1} & $\dots$ & $\dots$\\
\hline
\end{tabular}
\hspace{6pt}$\leadsto$\hspace{3pt}
 \begin{tikzpicture}[overlay, remember picture, yshift=.25\baselineskip, shorten >=.5pt, shorten <=.5pt]
  \draw [->] ([yshift=3pt]{pic cs:X9'})  [line width=0.2mm] to ([yshift=3pt]{pic cs:Y9});
	\draw [->] ([yshift=3pt]{pic cs:Y9'})  [line width=0.2mm] to ([yshift=3pt]{pic cs:Z9});
  \draw ([yshift=7pt]{pic cs:X9'})  edge[line width=0.2mm, out=25,in=135,->] ([yshift=6pt]{pic cs:Z9});
	\draw ([yshift=8pt]{pic cs:U9})  edge[line width=0.2mm, out=20,in=135,->] ([yshift=8pt]{pic cs:Z9});
  \end{tikzpicture}
\begin{tabular}{|c|c|c|c|}
\hline
 \multicolumn{4}{|l|}{ } \\
 \multicolumn{4}{|l|}{U\tikzmark{U10} \, \ \tikzmark{X10}X\tikzmark{X10'} \ \ \  \tikzmark{Y10}Y\tikzmark{Y10'} \ \, \  \tikzmark{Z10}Z} \\
\hline
 2 & 1 & \textbf{2} & $\dots$\\
\hline
 3 & 1 & \textbf{2} & $\dots$\\
\hline
 1 & 1 & \textbf{2} & $\dots$\\
\hline
 4 & 1 & \textbf{2} & $\dots$\\
\hline
\end{tabular}
\hspace{6pt}$\leadsto$\hspace{3pt}
 \begin{tikzpicture}[overlay, remember picture, yshift=.25\baselineskip, shorten >=.5pt, shorten <=.5pt]
  \draw [->] ([yshift=3pt]{pic cs:X10'})  [line width=0.2mm] to ([yshift=3pt]{pic cs:Y10});
	\draw [->] ([yshift=3pt]{pic cs:Y10'})  [line width=0.2mm] to ([yshift=3pt]{pic cs:Z10});
  \draw ([yshift=7pt]{pic cs:X10'})  edge[line width=0.2mm, out=25,in=135,->] ([yshift=6pt]{pic cs:Z10});
	\draw ([yshift=8pt]{pic cs:U10})  edge[line width=0.2mm, out=20,in=135,->] ([yshift=8pt]{pic cs:Z10});
  \end{tikzpicture}
\begin{tabular}{|c|c|c|c|}
\hline
 \multicolumn{4}{|l|}{ } \\
 \multicolumn{4}{|l|}{U\tikzmark{U11} \, \ \tikzmark{X11}X\tikzmark{X11'} \ \  \ \tikzmark{Y11}Y\tikzmark{Y11'} \, \; \; \; \ \tikzmark{Z11}Z} \\
\hline
 2 & 1 & 2 & \textbf{4}\\
\hline
 3 & 1 & 2 & \textbf{4}\\
\hline
 1 & 1 & 2 & \small$\hat f_Z(1,1,2)$\normalsize\\
\hline
 4 & 1 & 2 & \textbf{3} \\
\hline
\end{tabular}

 \begin{tikzpicture}[overlay, remember picture, yshift=.25\baselineskip, shorten >=.5pt, shorten <=.5pt]
  \draw [->] ([yshift=3pt]{pic cs:X11'})  [line width=0.2mm] to ([yshift=3pt]{pic cs:Y11});
	\draw [->] ([yshift=3pt]{pic cs:Y11'})  [line width=0.2mm] to ([yshift=3pt]{pic cs:Z11});
  \draw ([yshift=7pt]{pic cs:X11'})  edge[line width=0.2mm, out=25,in=150,->] ([yshift=6pt]{pic cs:Z11});
	\draw ([yshift=8pt]{pic cs:U11})  edge[line width=0.2mm, out=20,in=140,->] ([yshift=8pt]{pic cs:Z11});
  \end{tikzpicture}
\end{center}
In the first step, we changed the value of $X$ to $1$ in all rows. Next, the $Y$ column was filled using the fact that, according to the graph, $Y$ is determined by $X$; and observing that, in the initial team, rows that have value $1$ for $X$ have value $2$ for $Y$.
 Finally, we evaluated $Z$ (which could not have been done until we knew the values for $Y$); the procedure is composite. In the first and second row we obtained the value 4 for $Z$ as before, by checking, on the initial team, the rows that assume values $(2,1,2)$ (resp. $(3,1,2)$) over $U,X$ and $Y$. For the fourth row, we made use of the invariant functions: 
$\mathcal F_T(Z)(4,1,2)=3$.
 The value that $Z$ should assume in the third row cannot be reconstructed by looking at the initial team $T^-$, nor by using the information stored in $\mathcal F_T$; this can happen if the team is partially defined. Therefore, we insert a formal term to remind ourselves that the value for $Z$ in this row should be obtained applying an appropriate function $f_Z(U,X,Y)$ to the triple $(1,1,2)$ (if only we knew what what function it is).
We wrote $\hat f_Z$ as a formal symbol distinguished from the function $f_Z$.\footnote{More generally, complex terms with composition of many formal function symbols may be generated.}
Notice now that we have no uncertainties about the $Y$ column; so, it is natural to state that $T_{X=1}\models Y=2$, and that, therefore, $T\models X=1 \cf Y=2$. 

\end{example}

One must be careful in working out the details of the algorithm which
constitutes an intervention. The order of the updates of the descendents
of $X$ turns out not to be trivial, and it is not clear a priori whether
the algorithm will terminate. In case the causal team is partially defined, as in the example, there is also the problem that the information encoded in the
causal team may turn out to be insufficient for generating, under
intervention, a proper causal team, and thereby we must admit teams
which assign formal terms to some variables.

We begin considering the simplest case of \emph{recursive}, \emph{fully defined} causal teams. To take care of the order of the updating of variables, we introduce a notion of distance between (sets of) variables.

\begin{df}
Given a graph $G=(\SET V,E)$ and $\SET{X}\subseteq \SET V$, 

\begin{itemize}
\item We denote as $G_{-\SET{X}}=(\SET V,E_{-\SET{X}})$ the graph obtained by removing all arrows going into some vertex of $\SET X$ (i.e., an edge $(V_1,V_2)$ is in $E_{-\SET{X}}$ iff it is in $E$ and $V_2\notin \SET X$). Notice that, in the special case that $\SET{X} = \{X\}$, the set of directed paths of $G_{-\SET{X}}$ starting from $X$ coincides with the set of directed paths of $G$ starting from $X$.

\item Let $Y\in \SET V$. We call \textbf{(evaluation) distance} between $\mathbf{X}$ and $Y$ the value $d_G(\mathbf{X},Y)= sup\{length(P) |$ $ P \text{  directed path of $G_{-\SET{X}}$ going from some  $X\in \mathbf{X}$} \text{ to } Y\}$. In case no such path exists, $d_G(\mathbf{X},Y):=-1$. Clearly, if the graph is finite and acyclic, $d_G(\mathbf{X},Y)\in \mathbb{N}\cup\{-1\}$ for any pair $\mathbf{X},Y$. 
\end{itemize}
\end{df}

We write $\SET X = \SET x$ as an abbreviation for a conjunction of the form $X_1=x_1\land\dots \land X_n=x_n$. Let $\SET X = \SET x$ be a \emph{consistent} conjunction (that is, it does not contain two conjuncts of the form $X=x$ and $X=x'$, for $x\neq x'$). Then, applying the algorithm $do(\SET X = \SET x)$ to a recursive, fully defined causal team $T$ amounts to:\\

Stage $0$. Delete all arrows coming into $\SET X$, and replace each assignment $s\in T$ with $s(\SET{x}/\SET{X})$. Denote the resulting team\footnote{A warning: the teams $T_n$ produced before the last step of the algorithm may fail to form a causal team together with the other components described, because of violations of conditions b) and c) of the definition of causal team.} as $T_0$. Replace $\mathcal F_T$ with its restriction $\mathcal F_T'$ to $\SET V\setminus\SET X$. \\

Stage $n+1$. If $\{Z_1,\dots, Z_{k_{n+1}}\}$ is the set of all the variables $Z_j$ such that $d_{G(T)}(\SET{X},Z_j) = n+1$, define a team $T_{n+1}$ by replacing each $s\in T_n$ with the assignment $s(f_{Z_1}(s(PA_{Z_1}))/Z_1, \dots, f_{Z_{k_{n+1}}}(s(PA_{Z_{k_{n+1}}}))/Z_{k_{n+1}})$. \\ 

End the procedure after step $\hat n = sup\{d_{G(T)}(\SET{X},Z)|Z\in dom(T)\}$.\\

In case the intervention $do(\SET{X}=\SET{x})$ is a terminating algorithm on $T$, we define the causal team $T_{\SET{X}=\SET{x}}$ (of endogenous variables $\SET V\setminus \SET X$) as the quadruple $(T^{\hat n},G(T)_{-\SET X},\mathcal R_T, \mathcal F_T')$ which is produced when $do(\SET{X}=\SET{x})$ is applied to $T$. It is straightforward to prove (even in case the causal team has infinite ranges for some variables) that 

\begin{teo}
If $G(T)$ is a finite acyclic graph, then $T_{\SET{X}=\SET{x}}$ is well-defined.
\end{teo}


Furthermore, our definition of intervention has a kind of internal consistency: applying $do(\SET X=\SET x)$ is the same as sequentially applying interventions of the form $do(X=x)$, for each conjunct $X=x$ of $\SET X=\SET x$, in any order. This statement is a special case of the following two results:

\begin{teo}\label{IMPEXP}
Let $T$ be a recursive causal team, $\mathbf{X},\mathbf{Y}\in dom(T)$ such that $\mathbf{X} \cap \mathbf{Y}= \emptyset$, $\mathbf{x}\in Ran(\mathbf{X})$, and $\mathbf{y}\in Ran(\mathbf{Y})$. Then $T_{\mathbf{X}=\mathbf{x}\land \mathbf{Y}=\mathbf{y}} = (T_{\mathbf{X}=\mathbf{x}})_{\mathbf{Y}=\mathbf{y}}$.
\end{teo}

\begin{teo}\label{PERM}
Let $T$ be a recursive causal team, $\mathbf{X},\mathbf{Y}\in dom(T)$ such that $\mathbf{X} \cap \mathbf{Y}= \emptyset$, $\mathbf{x}\in ran(\mathbf{X})$, and $\mathbf{y}\in ran(\mathbf{Y})$. Then $(T_{\mathbf{X}=\mathbf{x}})_{\mathbf{Y}=\mathbf{y}} = (T_{\mathbf{Y}=\mathbf{y}})_{\mathbf{X}=\mathbf{x}}$.
\end{teo}

The first of these two theorems is proved by a somewhat complex double induction argument on the distances $d(\SET X,Z)$ and $d(\SET Y,Z)$ (for each variable $Z$). See \cite{BarSan2017} for details. The second theorem follows easily from the first: under the hypotheses, Theorem \ref{IMPEXP} entails the equalities $(T_{\mathbf{X}=\mathbf{x}})_{\mathbf{Y}=\mathbf{y}} = T_{\mathbf{X}=\mathbf{x}\land \mathbf{Y}=\mathbf{y}}$ and $(T_{\mathbf{Y}=\mathbf{y}})_{\mathbf{X}=\mathbf{x}} = T_{\mathbf{Y}=\mathbf{y}\land\mathbf{X}=\mathbf{x} }$; but since the order of variables is irrelevant in the definition of the $do$ algorithm, we also have $T_{\mathbf{X}=\mathbf{x}\land \mathbf{Y}=\mathbf{y}} = T_{\mathbf{Y}=\mathbf{y}\land \mathbf{X}=\mathbf{x}}$; transitivity yields the desired result. 

Having defined the intervened team $T_{\SET X=\SET x}$, we are immediately led to a semantical clause for counterfactuals of the form $\SET X=\SET x \cf \psi$:
\[
T\models \SET X=\SET x \cf \psi \iff T_{\SET X=\SET x} \models \psi.
\]
In case the antecedent is inconsistent (i.e., it contains two conjuncts $X_i = x_i, X_i = x_i'$ with $x_i \neq x_i'$), the corresponding intervention is not defined; in this case, we postulate the counterfactual to be true.

\section{The logic of recursive, fully defined causal teams} \label{SECRECPAR}

We call the (basic) \emph{language of causal dependence}, $\mathcal{CD}$, the language formed by the following rules:
\[
Y = y \ | \ Y\neq y \ | \ \dep{\SET X}{Y} \ | \ \psi \land \chi \ | \ \psi \land \chi \ | \ \theta \supset \chi \ | \ \SET X= \SET x \cf \chi
\]
for $Y,\SET{X}$ variables, $y,\SET x$ values, $\psi,\chi$ formulae of $\CD$, $\theta$ formula of $\CD$ without dependence atoms. The semantics for this language, evaluated over recursive fully defined causal teams, is given by the clauses presented in earlier sections. We also call $\CO$ (the \emph{causal-observational} language) the fragment of $\CD$ which lacks dependence atoms. We consider also an extension $\CO^{neg}$ of $\CO$ with a ``dual'' negation, whose semantics is defined by:

\begin{itemize}
\item $T\models \neg \psi$ iff for all $s\in T^-$, $\{s\}\not \models \psi$.\footnote{This atypical formulation of dual negation is justified by the flatness of the language $\CO$, entailed by Theorem \ref{TEOFLAT}.}
\end{itemize}

(Here, and in the following, we abuse notation and write $\{s\}$ for the \emph{causal} subteam $S$ of $T$ whose support $S^-$ is the singleton team $\{s\}$.)

In this section, we will give a short overview of the logical properties of these languages; we refer the reader to the preprint \cite{BarSan2017} for a more detailed account. First of all, we underline some global properties:

\begin{teo}
The logic $\mathcal{CD}$ is downwards closed, that is: if $\varphi\in \mathcal{CD}$, $T$ is a recursive\footnote{This statement (as the next one) holds more generally for fully defined causal teams with at most unique solution, to be introduced in a later section.} fully defined causal team, $T'$ is a causal subteam of $T$, and $T\models\varphi$, then also $T'\models\varphi$.
\end{teo}

\begin{teo}
The logic $\mathcal{CD}$ has the empty team property, that is: for every recursive, fully defined causal team $T$ with support $T^- = \emptyset$, and every $\varphi\in \mathcal{CD}$ with variables in $dom(T)$, $T\models\varphi$.
\end{teo}

\begin{teo} \label{TEOFLAT}
The logic $\CO^{neg}$ is flat, that is: for every formula $\varphi$ of $\CO^{neg}$ and every recursive, fully defined causal team $T$, $T\models \varphi$ iff $\{s\}\models \varphi$ for every assignment $s\in T^-$.
\end{teo}

This last result shows that our approach is in a sense a ``conservative extension'' of the structural equation modeling approach: as long as the language is poor enough, the semantics of causal teams can be reduced to that of deterministic structural equation models (which can be identified with causal teams with singleton support). However, in presence of other operators (e.g. dependence atoms, or the probabilistic atoms and boolean  disjunction that will be considered in the following sections) the use of causal teams is essential.

The proofs of these three theorems are routine inductions on the syntax of formulas. However, we wish to point out that the third theorem makes an essential use of the following fact: by applying an intervention to a causal team whose support is a singleton set, one obtains again a causal team with singleton support. This is a property which is guaranteed for recursive causal teams, or, more generally, for fully defined causal teams with unique solutions (see next section). 

Let us consider some further logical features of our framework. 
Unlike in the structural equation framework, the stronger variant of the law of excluded middle
\[
(S-EM): \hspace{15pt}\text{For any team } T, T\models \psi \text{ or } T\models \neg \psi
\] 
fails.
 Here is a very simple counterexample to it; the team

\begin{center}
\begin{tabular}{|c|}
\hline
 X  \\
\hline
  1 \\
\hline
  2 \\
\hline
\end{tabular}
\end{center}
does not satisfy $X=1$ nor its negation $X\neq 1$. A similar example shows that the following  strong form of the law of \emph{conditional} excluded middle
\[
(S-CEM): \text{ For every causal team $T$, } T\models \theta\cf\chi \text{ or } T\models \theta\cf\neg\chi
\]
fails. Within $\CO^{neg}$, however, the internalized versions of these laws (i.e. the statements that, for all recursive fully defined teams $T$, $T\models \psi \lor \neg\psi$, resp. $T\models \chi \cf (\psi \lor \neg\psi)$) are valid, due to flatness.  

Three laws that are often considered in relation to natural language and Lewis-Stalnaker counterfactuals (see e.g. \cite{Sid2010}) are the so-called \emph{importation, exportation} and \emph{permutation} laws; there are counterexamples for them in both contexts. Two results mentioned before (Theorems \ref{IMPEXP} and \ref{PERM})  provide sufficient conditions for the validity of these laws; their assumptions can be further relaxed as follows: assuming that \emph{the conjunction $\SET X = \SET x\land \SET Y = \SET y$ is consistent}, the following rules of inference 
\[
(IMP): \frac{\SET X=\SET x \cf (\SET Y=\SET y \cf \chi)}{(\SET X=\SET x \land \SET Y=\SET y) \cf \chi}  \hspace{25pt} (EXP): \frac{(\SET X=\SET x \land \SET Y=\SET y) \cf \chi}{\SET X=\SET x \cf (\SET Y=\SET y \cf \chi)}
\]
\[
(PERM): \frac{\SET X=\SET x \cf (\SET Y=\SET y \cf \chi)}{\SET Y=\SET y \cf (\SET X=\SET x \cf \chi)}
\]
are sound. More generally, the following ``overwriting'' rule (similar to an axiom discovered in \cite{Bri2012}) can be applied also in case $\SET X = \SET x\land \SET Y = \SET y$ is inconsistent:
\[
(CF-OUT): \frac{\SET{X}=\SET{x} \cf (\SET{Y}=\SET{y} \cf \psi)}{(\SET{X'}=\SET{x'} \land \SET{Y}=\SET{y}) \cf \psi};
\]
here $\SET{X'}=\SET{x'}$ is a conjunction of all the atoms of $\SET{X}=\SET{x}$ that contain no occurrences of variables from $\SET Y$. 

Galles\&Pearl (\cite{GalPea1998}) and Halpern (\cite{Hal2000}) provide an axiom system for (a case slightly more general than) recursive structural equation models. Their system can be adapted to our language $\CO^{neg}$ using the trick of transforming material implications into rules of inference. The resulting system (see \cite{BarSan2017}) is sound. However, $\CO^{neg}$ is more general than Halpern's language in that we allow counterfactuals and selective implications to occur in the consequents of counterfactuals\footnote{This has important consequences, such as the failure of modus ponens for $\cf$, and the failure of a version of Lewis's weak centering axiom. See also \cite{Bri2012}.}. Therefore, in order to obtain a completeness result, we need extra rules in order to extract these kinds of implications from consequents, or, vice versa, in order to insert them into consequents. The elimination and introduction of consequents can be performed by using the overwriting rule CF-OUT and an appropriate inverse, in case this consequent is a counterfactual statement; in case it is a selective implication, one can use the rule
\[
(SEL-OUT): \frac{\SET X = \SET x\cf(\psi \supset \chi)}{(\SET X = \SET x\cf\psi) \supset (\SET X = \SET x \cf\chi)}
\]
and its inverse. Similar extraction and introduction rules are available for the connectives $\land$, $\lor$ and $\neg$.

\section{Interventions on more general classes of causal teams} \label{SECMOREGEN}

We consider here the possibility of extending the notion of intervention on a causal team beyond the recursive, fully defined case.

\subsection{The (recursive) partially defined case}

In the case of a \emph{recursive, partially defined} team $T$, we first transform $T$ into an appropriate fully defined team $T'$, and then apply the algorithm from section \ref{SECINTERV} to $T'$. In order to define $T'$, we need first of all to extend the ranges of the variables of $T$ by allowing them to take as values also formal terms, as in example \ref{EXINTERV}. We call $L_{G(T)}$ the set of function symbols $\hat f_X$ (of arity $card(PA_X)$), for each endogenous variable $X$. We call $G(T)$-\emph{terms} the terms generated from variables in $dom(T)$ and from symbols in $L_{G(T)}$ by the obvious inductive rules; the set of all $G(T)$-terms will be denoted as $Term_{G(T)}$. 
 We then define the range component of $T'$ by: $\mathcal{R}_{T'}(X)=\mathcal{R}_T(X) \cup Term_{G(T)}$\footnote{Actually, only a finite number of formal terms are needed in each intervention. It is therefore possible to give a more  restrictive definition which preserves the finiteness of variable ranges.} 
for each $X\in G(T)$. 

Secondly, for $T'$ to be fully defined, we need the domains of the invariant functions to coincide with the ranges of the parent variables ($dom(f_X)=Ran(PA_X)$). Therefore, we have to redefine each $\mathcal F_T(X)$ component over the whole range $\mathcal R_T(PA_X)$. Let $pa_X \in \mathcal R_{T'}(X)$ be a sequence of values for $PA_X$. There are three possible cases: 1) $pa_X\in dom(\mathcal F_T(X))$; in this case we keep $\mathcal F_{T'}(X)(pa_X):=\mathcal F_{T}(X)(pa_X)$. Otherwise, 2) $pa_X\notin dom(\mathcal F_T(X))$, but there is an assignment $s\in T^-$ such that $s(PA_X)=pa_X$; in this case we set $\mathcal F_{T'}(X)(pa_X):= s(X)$ (i.e., we transfer information from the team component $T^-$ to the function component $\mathcal F_{T'}$). Otherwise, 3) we define  $\mathcal F_{T'}(X)(pa_X)$ to be the formal term $\hat f_X(pa_X)$. (Cf. example \ref{EXINTERV} for a justification of the three cases).

At this point, the algorithm $do(\SET X = \SET x)$ described in section \ref{SECINTERV} can be applied, and it will produce a causal team, some of whose entries may consist of formal terms. In the next section we will sketch  some ideas for the usage of these causal teams  as semantical objects for formal languages.

\subsection{The (fully defined) nonrecursive case}

In case a causal team is not recursive (i.e., its graph is cyclic), the algorithm above may well fail to terminate. However, if the causal team is fully defined and satisfies some further constraints, we can still find reasonable (but not necessarily computable) notions of intervention. One such constraint was isolated by Galles\&Pearl (\cite{GalPea1998}): they considered the case of systems of  structural equations \emph{with unique solutions}, defined as follows: 1) for fixed values of the exogenous variables, the system has a unique solution, and 2) each ``intervened'' system of equations obtained from the initial one by replacing some equations of the form $X:=f_X(PA_X)$ with constant equations $X:=x$ still has a unique solution for each choice of values for the exogenous variables. Since causal teams encode in an obvious way a system of modifiable structural equations, we can as well define causal teams with unique solutions. In this case, the natural way to define an intervention $do(X=x)$ on the team is to replace each assignment $s\in T^-$ with the (unique) assignment $t$ which encodes the solution of the intervened system for the choice $s(\SET{U})$ of values for the exogenous variables\footnote{In case the intervention acts also on some of the exogenous variables, this idea should be modified in an obvious way.}. The definition of the other components of the causal team produced by the intervention is straightforward. We do not expect any significant differences in the logical features of (fully defined) nonrecursive causal teams with unique solutions in comparison to  their recursive relatives. 


Analogous definitions could be given of causal teams with \emph{at most unique solutions} and of interventions over them (the idea being that, whenever a modified structural equation system admits no solution for $s(\SET U)$, the assignment $s$ should be discarded). We expect the corresponding logic to differ significantly from the case of unique solutions.

The general nonrecursive, fully defined case, where multiple solutions are allowed, is problematic. One might choose to add, to the intervened team, all the assignments that correspond to solutions of the modified equations. However, there seems to be no general criterion for deciding whether all such solutions should be given equal probabilistic weight (see next sections); this reflects general problems in the interpretation of nonrecursive causal systems (\cite{StrWol1960}). A second option might be to model such an intervention as producing not one, but multiple teams, corresponding to possible different outcomes of the intervention. This set of ``accessible teams'' would then induce a nontrivial modality, making it reasonable to treat counterfactuals as necessity operators in a dynamic logic setting (in the spirit of \cite{Hal2000}).




\section{Falsifiability and admissibility} \label{SECFALADM}



Interventions, when applied to a (recursive) partially defined causal team, can generate teams with formal entries. How should we evaluate statements which involve variables whose columns are not filled with proper values? 
Usually, we cannot ascertain their truth; e.g., we cannot assert $Y=3$ in a team whose non-formal entries for $Y$ are all equal to $3$. Yet, in some cases we might be able to observe the falsity of such statements; i.e., to state their contradictory negation. Let us write $\downarrow s(X)$ to signify that $s(X)$ is a value, not a formal term. Let $T$ be a team, possibly with formal entries. We read $T\models^f \psi$ as ``$\psi$ is falsifiable in $T$''.
We propose the clauses:
\begin{itemize}
\item $T\models^f X=x \text{ (resp. $X\neq x$)}\text{ if there is } s\in T^- \text{ such that } \downarrow s(X) \text{ and } s(X) \neq x$ \text{ (resp. $s(X) = x$)}
\item $T\models ^f \dep{\SET X}{Y} \text{ if there are } s,s'\in T^- \text{ such that } s(\SET X) = s'(\SET X), \downarrow s(Y),\downarrow s'(Y) \text{ and } s(Y) \neq s'(Y)$ 
\item $T\models^f \psi\land \chi$ if $T\models^f \psi$ or $T\models^f \chi$ 
\item $T\models^f \psi\lor \chi$ if for all subteams $T_1,T_2$ of $T$ with $T_1^-\cup T_2^-=T^-$, we have $T_1 \models^f \psi$ or $T_2\models^f \chi$
\item $T\models^f \SET X = \SET x \cf \chi$ if $T_{\SET X = \SET x} \models^f \chi$.
\end{itemize}

Coming up with a clause  for selective implication is less straightforward; we propose the following.  Given $\psi$ $\CO$ formula, let $\SET{V}^\psi$ be the set of variables occurring in $\psi$; define $T^\psi_*:=T^\psi \cup \{s\in T^- | \not\downarrow s(V) \text{ for some } V\in\SET{V}^\psi\}$. Then:
\begin{itemize}
\item $T\models^f \psi \supset \chi$ if $T^\psi_* \models^f \chi$
\end{itemize} 
As a justification for this clause, consider the team

\begin{center}
\begin{tabular}{|c|c|}
\hline
 \multicolumn{2}{|c|}{$\hspace{-5pt}X \ \rightarrow  Y$} \\
\hline
 2 & 1 \\
\hline
 1 & $\hat f_Y(1)$ \\
\hline
\end{tabular}
\end{center}

It seems unreasonable to assert that this team falsifies the formula $Y=1 \supset X=2$, because, as long as we do not have full knowledge of the function $f_Y$, we cannot decide  whether $\hat f_Y(1)$ is meant to denote $1$ or some other number; therefore, we do not know whether the second assignment is compatible or not with our selection - if it were, then the formula would be falsified, otherwise it would not be. We opt for the more cautious alternative.

We might also want to assert that some proposition is \emph{admissible} in the team, that is, consistent with the data we possess. The following seem to be reasonable clauses for the atomic formulas: 

\begin{itemize}
 \item $T\models^a X=x$ (resp. $X\neq x$) if for all $s\in T^-$ such that $\downarrow s(X), s(X) = x$ (resp. $s(X) \neq x$)
\item $T\models^a \dep{\SET X}{Y} \text{ if for all } s,s'\in T^- \text{ such that }  \downarrow s(Y), \downarrow s'(Y), s(\SET X) = s'(\SET X)$, we have $s(Y) =s'(Y)$. 
\end{itemize}

We do not consider the general case; but we still give clauses for ``classical'' formulas in disjunctive normal form:

\begin{itemize}
\item $T\models^a \bigvee_{i=1..m} \bigwedge_{j=1..n(i)} P^i_j$ ($P^i_j$ being of the form $X^i_j = x^i_j$ or $X^i_j \neq x^i_j$) if there are subteams $T_i$ of $T$, for $i=1..m$, such that 
		\begin{enumerate}
		\item	$T_i \models^a  P^i_j$, for all $j = 1..n(i)$. 
		\item for each $j,j'=1..n(i)$, if $j\neq j'$, $P^i_j$ is $X^i_j = a$ and $P^i_{j'}$ is $X^i_{j'} = b$ (with $a\neq b$), then for all $s\in T_i^-$, $s(X^i_j) \neq s(X^i_{j'})$.
		\item for each $j,j'=1..n(i)$, if $j\neq j'$, $P^i_j$ is $X^i_j = a$ and $P^i_{j'}$ is $X^i_{j'} \neq a$, then for all $s\in T_i^-$, $s(X^i_j) \neq s(X^i_{j'})$.
\end{enumerate}	
\end{itemize}

The clauses 2. and 3. above refer to formal inequality between terms. To have an idea of the intuition behind clause 2., the reader may think, for example, of the problem of checking the admissibility of $X=1 \land Y=2$; imagine that there is a row in which both the $X$-column and the $Y$-columm contain the formal term $f(3, g(2))$; then, surely, the formula is not admissible (for $X=1 \land Y=2$ to hold in the team, it is necessary that the $X$ and $Y$-column differ on each row). Clause 3. has a similar rationale.

If we restrict attention to causal teams that are generated by interventions applied to causal teams without formal entries, clause 2. and 3. can be omitted, because in this case the same formal term cannot occur in distinct columns of the intervened causal team (since, say, all formal terms in the $X$-column are of the form $\hat f_X(\dots)$, while all formal terms in the $Y$-column are of the form $\hat f_Y(\dots)$).

\section{Introducing probabilities} \label{SECPROB}

Probabilistic notions of causation have been extensively studied in
the literature. Bayesian networks formulate causal relations in terms
of conditional probabilities on (typically acyclic) graphs enriched
with a joint probability distribution over the variables of the graph
(Pearl \cite{Pea2000}, Spirtes, Glymour and Scheines \cite{SpiGlySch1993}). Woodward also considers 
interventions on a variable that cause changes in the probability of another variable. 
In the context of team semantics, probabilities have been recently introduced via the notion of multiteam. A multiteam differs from a team in that it may feature multiple copies of the same assignment; it is therefore closer to a collection of experimental data than teams are. There have been at least two different approaches to the formalization of multiteams in the literature (\cite{Vaa2017}, \cite{DurHanKonMeiVir2016}). For simplicity, we simulate multiteams by means of teams. This can be accomplished by assuming that each team has an extra variable Key (never mentioned in the object languages) which assumes distinct values on distinct assignments of the same team. In this way, we can have two assignments that agree on all the significant variables and just differ on Key.
With this assumption, the definition of \emph{causal multiteam} can follow word by word the definition of causal team.

If we wish to talk about probabilities, it is natural to allow for more atomic formulas.
\begin{df}
The set of probabilistic literals is given by:
\[
\sim\hspace{-2pt}\alpha \ | \ Pr(\chi) \leq \epsilon \ | \ Pr(\chi) \geq \epsilon \ | \ Pr(\chi) \leq Pr(\theta) \ | \ Pr(\chi) \geq Pr(\theta)
\]
where $\alpha$ is a probabilistic literal, $\chi,\theta$ are formulas of $\CO$ and $\epsilon \in \mathbb{R}\cap [0,1]$. Literals and probabilistic literals without negation will be called atomic formulas.

The \emph{(basic) probabilistic causal language} ($\PCD$) is given by the following clauses:
\[
\alpha \ | \ \psi \land \chi \ | \ \psi \lor \chi \ | \ \psi \sqcup \chi \ | \ \theta\supset \psi \ | \ \SET X = \SET x\cf \psi
\]
where $\alpha$ is a literal or probabilistic literal, $\psi,\chi$ are $\PCD$ formulas, and $\theta$ a $\CO$ formula.
\end{df}

The symbols $\sim$ stands for contradictory negation ($T\models\sim\hspace{-2pt}\psi$ iff $T\not\models\psi$). We will use abbreviations such as $Pr(\chi) = \epsilon$ for $Pr(\chi) \leq \epsilon \land Pr(\chi) \geq \epsilon$, or $Pr(\chi) < \epsilon$ for $Pr(\chi) \leq \epsilon \land \sim\hspace{-2pt}Pr(\chi) \geq \epsilon$. The additional connective $\sqcup$ is known as boolean disjunction and its interpretation is given by the clause:
\begin{itemize}
\item $T\models \psi \sqcup \chi \iff T\models \psi$ or $T\models \chi$.
\end{itemize}
The statement ``either $X=x$ has probability less than one third, or greater than two thirds'' should be expressed as $Pr(X=x) <1/3 \sqcup Pr(X=x) > 2/3$, and not by means of  the earlier disjunction $\lor$. The reader can verify this point as soon as we give semantical clauses for the probabilistic literals. 
 For any $\CO$ formula $\chi$ and any causal team $T$ with nonempty finite support $T^-$, define the \textbf{probability of $\chi$ in $T$} as:
\[
Pr_T(\chi):= \frac{card(\{s\in T^- | \{s\}\models \chi\})}{card(T^-)}.
\]
It can be verified that this definition induces a probabilistic space over the subteams of $T^-$ that are definable by some $\CO$ formula. 

The semantics of probabilistic atoms can then be defined as:
\begin{itemize}
\item $T\models Pr(\chi)\leq \epsilon  \iff T^-\neq\emptyset \text{ and } Pr_T(\chi)\leq \epsilon$
\item $T\models Pr(\chi)\leq Pr(\theta) \iff T^-\neq\emptyset \text{ and } Pr_T(\chi)\leq Pr_T(\theta)$
\end{itemize}
et cetera\footnote{Notice that, by definition, causal teams with empty support do \emph{not} satisfy the probabilistic atoms.}. It is easy to see that such a logic is \emph{not} downward closed; for example, a team such that less than half assignments satisfy $\chi$ will satisfy $Pr(\chi)\leq \frac{1}{2}$; but the subteam $T^\chi$ constituted \emph{only} of the assignments that satisfy $\chi$ will not satisfy $Pr(\chi)\leq \frac{1}{2}$. 

Can we define conditional probabilities in this kind of framework? Given two $\CO$ formulas $\chi_1$ and $\chi_2$, 
we write $Pr(\chi_2|\chi_1)\leq \epsilon$ as an abbreviation for $\chi_1 \supset Pr(\chi_2)\leq \epsilon$. Here is a proof that the abbreviation has the intended meaning: assuming $T^-\neq \emptyset$, 
\[
T\models \chi_1 \supset Pr(\chi_2)\leq \epsilon \iff
 T^{\chi_1}\models Pr(\chi_2)\leq \epsilon \iff
 \frac{card(\{s\in (T^{\chi_1})^- | \{s\}\models \chi_2\})}{card((T^{\chi_1})^-)}\leq \epsilon
\]
\[
\iff \frac{card(\{s\in (T^{\chi_1})^- | \{s\}\models \chi_2\})}{card(T^-)}\frac{card(T^-)}{card((T^{\chi_1})^-)} \leq \epsilon
\iff \frac{Pr_T(\chi_1 \land \chi_2)}{Pr_T(\chi_1)} \leq \epsilon,
\]
and we observe that the left member in this last equation is the usual definition of the conditional probability $Pr_T(\chi_2|\chi_1)$. In case $T^- = \emptyset$, it is easily proved, instead, that $T\not\models \chi_1 \supset Pr(\chi_2)\leq \epsilon$.  Things work analogously for inequalities in the opposite direction, and for atoms of the form $Pr(\chi)\leq Pr(\theta)$. 
 
In the literature (e.g. \cite{Pea2000}) one finds ad hoc notations that mix interventions and probabilities; for example, $P(y|do(x),z) = \epsilon$ is used for a probability which is conditional on the outcome of an intervention (\emph{post-intervention} conditioning); the notation $P(Y_x|z) = \epsilon$ is used for the probability of a variable after the intervention, conditioned on \emph{pre-intervention} observations. These two cases are expressed, in $\PCD$, as $X=x \cf (Z=z \supset Pr(Y=y) = \epsilon)$, resp. $Z=z \supset (X=x \cf Pr(Y=y) = \epsilon)$; their difference amounts to a swap in the order of application of $\cf$ and $\supset$. Our formalism immediately shows that more varied possibilities could be considered, such as conditioning simultaneously pre- and post-intervention ($W=w \supset (X=x \cf (Z=z \supset Pr(Y=y) = \epsilon))$) or between two interventions $X=x \cf (Z=z \supset (W=w \cf Pr(Y=y) = \epsilon))$.

\section{Direct and total cause} \label{SECDIRTOT}

We show that the basic type-causal notions from Woodward (\cite{Woo2003}), \emph{direct} and \emph{total} cause, can be expressed in our languages, over causal teams which are finite,  recursive and fully defined. Quoting from Woodward:
\begin{quote}
A necessary and sufficient condition for $X$ to be a direct cause of $Y$ with respect to some variable set $\SET V$ is that there be a possible intervention on $X$ that will change $Y$ (or the probability distribution of $Y$) when all other variables in $\SET V$ besides $X$ and $Y$ are held fixed at some value by interventions. (\cite{Woo2003}, p.55)
\end{quote}

This definition is ambiguous in that it talks about a change in $Y$, but  does not say with respect to what the change is made; to $Y$'s actual value? To some possible value of $Y$, i.e., some $y\in Ran(Y)$? 
We resolve the ambiguity by stipulating that the values of $Y$ to be compared are generated by \emph{two} distinct interventions. 

 The kind of intervention that is needed in order to establish whether X is a direct cause of Y is an intervention on 
all variables in the domain except for $X$ and $Y$. For example, consider the causal team $T$ in the figure below
(with invariant functions $\mathcal{F}_Z(X):= X$ and $\mathcal{F}_Y(X,Y) := X+Y$). 
We show that $X$ is a direct cause of $Y$ in $T$. First of all we must fix all other variables (in this case, just $Z$) to an appropriate value (we choose $1$) by an intervention, which also removes the  arrow that enters in $Z$, and updates $Y$:

\begin{center}
$T$: \begin{tabular}{|c|c|c|}
\hline
 \multicolumn{3}{|c|}{ }\\ 
 \multicolumn{3}{|l|}{X\tikzmark{X14} \ \ \ \tikzmark{Y14}Z\tikzmark{Y14'} \ \  \tikzmark{Z14}Y} \\
\hline
 1 & 1 & 2 \\
\hline
 2 & 2 & 4 \\
\hline
 3 & 3 & 6 \\
\hline
\end{tabular}
 \begin{tikzpicture}[overlay, remember picture, yshift=.25\baselineskip, shorten >=.5pt, shorten <=.5pt]
  \draw [->] ([yshift=3pt]{pic cs:X14})  [line width=0.2mm] to ([yshift=3pt]{pic cs:Y14});
	\draw [->] ([yshift=3pt]{pic cs:Y14'})  [line width=0.2mm] to ([yshift=3pt]{pic cs:Z14});
  \draw ([yshift=7pt]{pic cs:X14})  edge[line width=0.2mm, out=55,in=125,->] ([yshift=7pt]{pic cs:Z14});
  \end{tikzpicture}
\hspace{10pt}$\leadsto$\hspace{10pt}
$T_{Z=1}$: \begin{tabular}{|c|c|c|}
\hline
 \multicolumn{3}{|c|}{ }\\ 
 \multicolumn{3}{|l|}{X\tikzmark{X17} \ \ \ \tikzmark{Y17}Z\tikzmark{Y17'} \ \ \tikzmark{Z17}Y} \\
\hline
 1 & \textbf{1} & \textbf{2} \\
\hline
 2 & \textbf{1} & \textbf{3} \\
\hline
 3 & \textbf{1} & \textbf{4} \\
\hline
\end{tabular}
 \begin{tikzpicture}[overlay, remember picture, yshift=.25\baselineskip, shorten >=.5pt, shorten <=.5pt]
	\draw [->] ([yshift=3pt]{pic cs:Y17'})  [line width=0.2mm] to ([yshift=3pt]{pic cs:Z17});
  \draw ([yshift=7pt]{pic cs:X17})  edge[line width=0.2mm, out=55,in=125,->] ([yshift=7pt]{pic  cs:Z17});
  \end{tikzpicture}
\end{center}

Then we intervene in two different ways on $X$, by $do(X=1)$ and $do(X=2)$:
\begin{center}
\begin{tabular}{|c|c|c|}
\hline
 \multicolumn{3}{|c|}{ }\\ 
 \multicolumn{3}{|l|}{X\tikzmark{X15} \ \ \ \tikzmark{Y15}Z\tikzmark{Y15'} \ \ \tikzmark{Z15}Y} \\
\hline
 \textbf{1} & 1 & \textbf{2} \\
\hline
\end{tabular}
 \begin{tikzpicture}[overlay, remember picture, yshift=.25\baselineskip, shorten >=.5pt, shorten <=.5pt]
	\draw [->] ([yshift=3pt]{pic cs:Y15'})  [line width=0.2mm] to ([yshift=3pt]{pic cs:Z15});
  \draw ([yshift=7pt]{pic cs:X15})  edge[line width=0.2mm, out=55,in=125,->] ([yshift=7pt]{pic cs:Z15});
  \end{tikzpicture}
\begin{tabular}{|c|c|c|}
\hline
 \multicolumn{3}{|c|}{ }\\ 
 \multicolumn{3}{|l|}{X\tikzmark{X16} \ \ \ \tikzmark{Y16}Z\tikzmark{Y16'} \ \ \tikzmark{Z16}Y} \\
\hline
  \textbf{2} & 1 & \textbf{3} \\
\hline
\end{tabular}
 \begin{tikzpicture}[overlay, remember picture, yshift=.25\baselineskip, shorten >=.5pt, shorten <=.5pt]
	\draw [->] ([yshift=3pt]{pic cs:Y16'})  [line width=0.2mm] to ([yshift=3pt]{pic cs:Z16});
  \draw ([yshift=7pt]{pic cs:X16})  edge[line width=0.2mm, out=55,in=125,->] ([yshift=7pt]{pic cs:Z16});
  \end{tikzpicture}	
\end{center}
The fact that the two interventions generate distinct values for $Y$ proves that $X$ is a direct cause of $Y$. The specific form of these kinds of interventions makes it so that, \emph{if there is an arrow from $X$ to $Y$}, the intervention enforces a team with constant columns;
that is, a singleton causal team is produced. 

Let $Fix(\SET z)$ be an abbreviation for $\bigwedge_{Z\in Dom(T)\setminus\{X,Y\}}Z=z$. Then, the fact that $X$ is a direct cause of $Y$ in $T$ can be expressed in $\CD$ as follows: $T\models DC(X;Y)$ iff 
\[
T\models\bigsqcup_{x\neq x', y\neq y',\SET z}[(Fix(\SET z) \land X=x) \cf Y=y] \land [(Fix(\SET z) \land X=x') \cf Y=y'].
\] 

In the probabilistic setting, applying the intervention described by $Fix(\SET z)$ does not in general shrink the multiteam to a singleton, because the resulting multiteam may still consist of multiple copies of one and the same assignment. Nevertheless, we can still define direct causation, $T\models PDC(X;Y)$:
\[
T\models \bigsqcup_{x\neq x',y,\SET z}[(Fix(\SET z) \land X=x) \cf Pr(Y=y) = 0] \land [(Fix(\SET z) \land X=x') \cf Pr(Y=y)=1].
\]
In a sense, we have a collapse of the probabilistic case to the deterministic one.


We now consider the notion of total cause, following again Woodward:
\begin{quote}
$X$ is a total cause of $Y$ if and only if there is a possible intervention on $X$ that will change $Y$ or the probability distribution of $Y$. (\cite{Woo2003}, p.51)
\end{quote}
Applying the kind of intervention described by Woodward, teams do not in general shrink to singletons. However, total cause can be equivalently defined as the existence of such interventions, to be applied after \emph{all nondescendants of $X$} have been fixed to some values. We denote by $Fix'(\SET w)$ the conjunction that expresses the intervention that fixes all nondescendants $\SET W$ of $X$ to $\SET w$. Such an intervention does shrink the causal team to a singleton, provided there is at least one directed path from $X$ to $Y$. We can thus express that $X$ is a total cause of $Y$ in $T$, $T\models TC(X;Y)$, by the clause:
\[
T\models \bigsqcup_{x\neq x', y\neq y',\SET w} Fix'(\SET w) \cf [(X=x\cf Y=y) \land (X=x'\cf Y=y')].
\]
A similar definition can be given in the probabilistic language, using the fact that only a finite number of distinct probability values can arise from a finite multiteam.

\bibliographystyle{eptcs}
\bibliography{iilogicsdoi}

\end{document}